%% file: main.tex
\pdfoutput=1

\PassOptionsToPackage{margin=0.89in}{geometry}
\PassOptionsToPackage{dvipsnames}{xcolor}
\documentclass[preprint,manuscript,10pt]{acmart}

\makeatletter
\def\mdseries@tt{m}
\makeatother

\setcopyright{none}
\acmDOI{}
\acmPrice{}

\input{tex/packages}
\input{tex/layout-tweaks}
\input{tex-common/macros}

\input{tex/macros}
\input{tex/listings}
\setcitestyle{nosort,numbers,comma}

\begin{document}
\title{The meaning of a program change is a change to the program's meaning\vspace*{-2ex}}
\input{tex/authors}

\twocolumn[\maketitle]

\input{page-1}
\input{page-2}
\onecolumn\newpage\twocolumn

\input{main.bbl}
\end{document}

%% file: tex/packages.tex
\usepackage{float}
\usepackage{mathpartir}
\usepackage{bm}
\usepackage{soul}
\usepackage{xspace}
\usepackage{mathtools}
\usepackage{hyperref}
\usepackage{breakurl}
\usepackage[scaled=0.7]{beramono}
\usepackage{pifont}
\usepackage{framed}
\usepackage{subdepth}
\usepackage{dashundergaps}
\usepackage{listings}
\usepackage{mleftright}
\usepackage{extarrows}
\usepackage{placeins}
\usepackage[skip=0pt]{caption}
\usepackage{xpatch}
\usepackage{tikz}
\input{tex/graph-common}

\usepackage[many]{tcolorbox}
\tcbuselibrary{listings}
\usepackage{flushend}

\input{tex-common/secbreak}

%% file: tex/graph-common.tex
\pgfdeclarelayer{values}
\pgfdeclarelayer{FV}
\pgfsetlayers{main,values,FV}

\tikzstyle{valueIn}=[rectangle,draw=black!50,anchor=west,xshift=2mm,minimum height=1.25em,text height=1.5ex,text depth=.25ex]
\tikzstyle{value}=[rectangle,draw=black!50,fill=black!20,anchor=west,xshift=2mm,minimum height=1.25em,text height=1.5ex,text depth=.25ex]
\tikzstyle{valueNew}=[value,thick,draw=ForestGreen,fill=ForestGreen!30,text=ForestGreen]
\tikzstyle{trace}=[rectangle,minimum height=1.25em,text height=1.5ex,text depth=.25ex]
\tikzstyle{traceNew}=[trace,text=ForestGreen]
\tikzstyle{subtrace}=[->,black!50]
\tikzstyle{subtraceNew}=[subtrace,ForestGreen!70]
\tikzstyle{subtraceChanged}=[subtrace,blue]
\tikzstyle{subvalue}=[->,black!50]
\tikzstyle{subvalueNew}=[->,thick,ForestGreen]
\tikzstyle{subvalueChanged}=[subvalue,blue,densely dotted,thick]
\tikzstyle{arg}=[->,draw=black]
\tikzstyle{argNew}=[->,draw=ForestGreen]
\tikzstyle{val}=[->,draw=black]
\tikzstyle{valNew}=[->,draw=ForestGreen]
\tikzstyle{nullVal}=[shape=rectangle,sharp corners,fill=black,inner sep=1pt,anchor=west]
\tikzstyle{every node}=[transform shape]
\tikzstyle{FV}=[fill=white!50,draw=red,anchor=south west,yshift=-1.25mm,xshift=-2mm,scale=.75,minimum height=4mm,minimum width=3mm,inner sep=0.5mm]

\tikzstyle{annotTrace}=[matrix,FV,start chain,inner sep=0mm,node distance=0mm,minimum height=0mm,text height=1.5ex,text depth=.25ex]
\tikzstyle{annotChild}=[on chain,scale=0.4875,inner sep=0.5mm,shape=circle,minimum height=0.25mm,minimum width=0.25mm,fill=red,text height=0ex,text depth=0ex]
\tikzstyle{annotText}=[on chain,scale=0.4875,inner sep=0.5mm,minimum height=4mm]
\tikzstyle{annotLink}=[red,thin,opacity=0.6]

%% file: tex/layout-tweaks.tex
\everymath{\displaystyle}

\allowdisplaybreaks

\hypersetup{ 
    colorlinks=false,
    pdfborder={0 0 0},
}

\makeatletter
\xpatchcmd
   {\ps@firstpagestyle}
   {Manuscript submitted to ACM}
   {\emph{Submitted to Incremental Computation 2019}}
   {\typeout{First patch succeeded}}
   {\typeout{first patch failed}}
\xpatchcmd
   {\ps@standardpagestyle}{Manuscript submitted to ACM}
   {\emph{Submitted to Incremental Computation 2019}}
   {\typeout{Second patch succeeded}}
   {\typeout{Second patch failed}}
\@ACM@manuscriptfalse
\makeatother

%% file: tex-common/macros.tex
\newcommand*{\ttt}[1]{\texttt{#1}}
\newcommand*{\kw}[1]{{\text{\tt{#1}}}} 

\DeclareTextFontCommand{\textbfit}{%
  \fontseries\bfdefault 
  \itshape
}

\DeclareSymbolFont{bbsymbol}{U}{bbold}{m}{n}
\DeclareMathSymbol{\bbsemi}{\mathbin}{bbsymbol}{"3B}

\newenvironment{nop}{}{}
\newenvironment{sdisplaymath}
   {\begin{nop}\small\begin{displaymath}}
   {\end{displaymath}\end{nop}\ignorespacesafterend}
\newenvironment{smathpar}
   {\begin{nop}\small\begin{mathpar}}
   {\end{mathpar}\end{nop}\ignorespacesafterend}

\newenvironment{mathfig}{\begin{sdisplaymath}}{\end{sdisplaymath}}

\makeatletter
\newbox\sf@box
  {\def\sf@one{#1}%
   \def\sf@two{#2}%
   \setbox\sf@box\hbox
      \bgroup}%
  { \egroup
   \ifx\@empty\sf@two\@empty\relax
     \def\sf@two{\@empty}
   \fi
   \ifx\@empty\sf@one\@empty\relax
      \subfloat[\sf@two]{\box\sf@box}%
   \else
      \subfloat[\sf@one][\sf@two]{\box\sf@box}%
   \fi}
\makeatother

\RequirePackage{xcolor}

\definecolor{highlightcolor}{rgb}{1.0,0.8,0.8}
\definecolor{shadecolor}{rgb}{0.9,0.9,0.9}
\definecolor{lightgray}{rgb}{0.8,0.8,0.8}
\newcommand*{\shadebox}[1]{\fcolorbox{lightgray}{shadecolor}{\raisebox{0pt}[0.60\baselineskip][0.05\baselineskip]{#1}}}

\makeatletter
\newcommand{\superimpose}[2]{%
  {\ooalign{$#1\@firstoftwo#2$\cr\hfil$#1\@secondoftwo#2$\hfil\cr}}}
\makeatother

\newsavebox{\vardisplaymathbox}

%% file: tex/macros.tex
\newcommand*{\New}[1]{$\textcolor{ForestGreen}{\kw{#1}}$}

\newcommand*{\eval}{\Rightarrow}
\newcommand*{\deltaeval}{\mathrel{\Rightarrow_{\delta}}}

\newcommand{\app}[1]{\kw{#1}} 

\newcommand*{\drawArg}[2]{\draw[arg] (#2) -- (#1);}
\newcommand*{\drawArgNew}[2]{\draw[argNew] (#2) -- (#1);}

\newcommand\myprime{\raise-0.2ex\hbox{\scalebox{0.8}{$\scriptstyle\prime$}}}
\newcommand{\oneprime}[1]{\smash{#1^{\myprime}}}

%% file: tex/listings.tex
\definecolor{verylightgray}{gray}{0.9}
\definecolor{lightgray}{gray}{0.5}
\definecolor{mediumgray}{gray}{0.45}
\newlength\lsthorizontalpadding
\newcommand*\lstnumberstyle{\ttfamily\scriptsize\textcolor{lightgray}}
\newlength\lstnumbersep
\newlength\lstnumberwidth
\lstdefinelanguage{code}{%
   morekeywords={match,as,fun}%
}

\newtcblisting{graylstlisting}[1][]
 {
  spartan,
  frame empty,
  boxsep=0mm,
  top=0pt,
  bottom=0pt,
  left=0pt,
  listing only,
  colback=verylightgray,
  listing options={
     numbersep=10pt%
    ,xleftmargin=0pt%
    ,aboveskip = 0pt%
    ,belowskip = 0pt%
    ,basicstyle=\linespread{0.9}\ttfamily%
    ,breaklines=true%
    ,tabsize=4%
    ,showstringspaces=false%
    ,numbers=left%
    ,numbersep=\lstnumbersep%
    ,numberstyle=\lstnumberstyle%
    ,framesep=0pt%
    ,xleftmargin=3pt
    ,framexleftmargin=\lsthorizontalpadding%
    ,xrightmargin=\lsthorizontalpadding%
    ,framexrightmargin=\lsthorizontalpadding%
    ,backgroundcolor=\color{verylightgray}%
    ,postbreak=\ding{229}\space%
    ,mathescape=true%
  },
 }

%% file: tex/authors.tex
\author{Roly Perera}
\orcid{0000-0001-9249-9862}
\affiliation{
   \institution{The Alan Turing Institute}
   \streetaddress{96 Euston Rd}
   \postcode{NW1 2DB}
   \city{London}
   \country{London, UK} 
}
\email{rperera@turing.ac.uk}

%% file: page-1.tex
\subsection*{Programming as dialogue}

\emph{Incremental} computations
\cite{paige77,teitelbaum81,acar05,mcsherry13,cai14} map input changes to output
changes, ideally faster than running from scratch on the new input. The program
itself is assumed to be fixed. It may be possible to replace functional inputs
with other functions, but changes to the program text itself are not usually
considered.

Programming is also an incremental endeavour \cite{lehman69,larman03}. The focus
here is on \emph{program} changes, often in the context of a fixed input such as
a test case. Since the entire goal is to bring about specific changes in
behaviour, the output changes that arise are of great interest to the
programmer. Live programming \cite{tanimoto90,hancock03,mcdirmid07} and
test-driven development \cite{beck02} recognise this, but neither provides a
fine-grained notion of change. New outputs appear wholesale, leaving it up to
the programmer to figure out what changed and why.

This suggests a research problem: developing an incremental, fine-grained
semantics which maps program changes to output changes, interpreting \emph{the
meaning of program changes} as \emph{changes to the program's meaning}.

As a toy example, suppose I wish to compute the sum of the squares of a list of
integers. I start with a simpler program, say a function which simply copies a
list:

\begin{graylstlisting}
f [] = []
f (x : xs) = x : f xs
\end{graylstlisting} %

\noindent As a sanity-check, I run my function at an argument:

\begin{graylstlisting}
f [1, 3] => [1, 3]
\end{graylstlisting}

Perhaps seeking a ``tabular'' view of the problem, I decide to pair each element
with its square, making the following edit (new code is shown in green):

\begin{graylstlisting}
f [] = []
f (x : xs) = $\New{(}$x, $\New{sq x)}$ : f xs
\end{graylstlisting}

\noindent I can now re-run the earlier test case, to check that the output has
become \kw{[(1, 1), (3, 9)]}. But what if the programming environment could
directly interpret my program change as an output change?

\begin{graylstlisting}
f [1, 3] => [$\New{(}$1$\New{, 1}$), $\New{(}$3$\New{, 9)}$]
\end{graylstlisting}

\noindent These are, after all, the intuitive consequences of editing the
program in the specific way that I did: my edit \emph{paired each element with
its square} (on the right). Here, for example, the output delta confirms my
intuition that, of the two occurrences of $\kw{1}$ in the revised output, it is
the second one which should be considered ``new''.

\vspace{3pt} \subsection*{Towards a semantics of program change}

Rather than considering in detail how this might this work, we content ourselves
with an informal exploration of the problem. Work on program slicing
\cite{perera12a,ricciotti17} and live programming with holes \cite{omar19}
assigns meaning to incomplete programs and their (partial) completions; this has
some of the flavour of what we want, but is a rather monotone notion. To support
structural reorganisations like moves and splices, we will suppose a \emph{term
graph} representation of programs \cite{plump99}, perhaps provided by a
\emph{structure editor} \cite{donzeau80} able to assign unique addresses to
subterms \cite{perera13}:

\vspace{3pt}
\input{fig/pair-sq-prog}
\\

The graphs on the left graphically represent the two defining clauses of \kw{f};
we place them side-by-side to establish them as disjoint parts of the same
definition. Inputs nodes (patterns) are shown in white, and output nodes
(values) in grey. The recursive call to $\kw{f}$ appears as an application node
with no output edge.

The program edit is shown graphically to the right, as a ``graph delta'' to the
cons clause of the definition. The new pair constructor and its child edges,
which should be read as pointers, are shown in green. The pointer to the first
element of the output is dotted blue, meaning the pointer value has changed;
this is necessarily the case here because it refers to a newly-created node.

What we seek, then, is a semantics that allows us to ``execute'' the edit to the
program to obtain a corresponding edit to its execution. Below left is the
original execution of the test case, represented as a term graph with a shape
resembling a big-step derivation:

\input{fig/pair-sq}
\\

\noindent To the right is the modified execution we would like to obtain, with
delta-colouring showing new nodes spliced into the computation in a way that
accounts for the output delta identified previously. What this illustrates is
that a suitable notion of execution delta should (we contend)
``operationalise'', or account for, output deltas, generalising the idea from
dynamic program slicing that parts of executions account for specific parts of
the output \cite{perera12a}.

In the other direction, we must seek an appropriate technical sense in which
those execution deltas ``correspond'' to the program deltas that gave rise to
them. As an example of a potential property of interest, one might consider
whether there exists a map from execution nodes to program nodes preserving the
graph structure and delta colouring.

%% file: fig/pair-sq-prog.tex
\begin{minipage}[t]{0.6\columnwidth}
\begin{center}
\begin{tikzpicture}[rounded corners=1pt,scale=0.7,baseline=(app-1)]

\node[trace] at (-2.5,0)
(app-nil-1)          {\app{f}};
\node[valueIn,left of=app-nil-1]
(in-nil-nil)      {\ttt{[]}};
\node[value,right of=app-nil-1]
(out-nil-nil)     {\ttt{[]}};

\drawArg{app-nil-1.west}{in-nil-nil.east}
\draw[val] (app-nil-1.east) -- (out-nil-nil.west);

\node[trace] at (1,0)
(app-1)          {\app{f}};
\node[valueIn,left of=app-1]
(in-cons-1)      {\ttt{:}};
\node[valueIn,below left of=in-cons-1]
(in-el-1)        {\ttt{x}};
\node[value,right of=app-1]
(out-cons-1)     {\ttt{:}};
\node[value,below left of=out-cons-1]
(out-el-1)       {\ttt{x}};

\drawArg{app-1.west}{in-cons-1.east}
\draw[subvalue] (in-cons-1.south west) -- (in-el-1.north east);
\draw[val] (app-1.east) -- (out-cons-1.west);
\draw[subvalue] (out-cons-1.south west) -- (out-el-1.north east);

\node[trace,below of=out-cons-1,xshift=-3mm,yshift=-4mm]
(app-2)          {\app{f}};
\node[valueIn,left of=app-2]
(in-cons-2)      {\ttt{xs}};

\draw[subvalue] (in-cons-1.south east) -- (in-cons-2.north west);
\drawArg{app-2.west}{in-cons-2.east}
\draw[subvalue] (out-cons-1.south) -- (app-2.north);

\draw[draw=black!40,densely dotted,rounded corners] (-3.85,-1.9) rectangle (2.65,0.5);
\draw[draw=black!40,densely dotted] (-0.75,-1.9) -- (-0.75,0.5);

\end{tikzpicture}
\end{center}
\end{minipage}%
\begin{minipage}[t]{0.35\columnwidth}
\begin{center}
\begin{tikzpicture}[rounded corners=1pt,scale=0.7,baseline=(in-cons-1)]

\node[valueIn] at (0,0)
(in-cons-1)      {\ttt{:}};
\node[trace,right of=in-cons-1]
(app-1)          {\app{f}};
\node[valueIn,below left of=in-cons-1]
(in-el-1)        {\ttt{x}};
\node[value,right of=app-1]
(out-cons-1)     {\ttt{:}};
\node[valueNew,below left of=out-cons-1,xshift=-2mm]
(out-el-1)       {\ttt{(,)}};
\node[value,below left of=out-el-1,yshift=-2mm]
(out-pairl-1)    {\ttt{x}};
\node[traceNew,right of=out-pairl-1]
(app-sq-1)       {\app{sq}};

\drawArg{app-1.west}{in-cons-1.east}
\draw[subvalue] (in-cons-1.south west) -- (in-el-1.north east);
\draw[val] (app-1.east) -- (out-cons-1.west);
\draw[subvalueChanged] (out-cons-1.south west) -- (out-el-1.north east);
\drawArgNew{app-sq-1.west}{out-pairl-1.east}
\draw[subvalueNew] (out-el-1.south) -- (out-pairl-1.north east);
\draw[subvalueNew] (out-el-1.south) -- (app-sq-1.north west);

\node[valueIn,below right of=in-cons-1,yshift=-16mm]
(in-cons-2)      {\ttt{xs}};
\node[trace,right of=in-cons-2]
(app-2)          {\app{f}};

\draw[subvalue] (in-cons-1.south east) -- (in-cons-2.north west);
\drawArg{app-2.west}{in-cons-2.east}
\draw[subvalue] (out-cons-1.south) -- (app-2.north);

\end{tikzpicture}
\end{center}
\end{minipage}

%% file: fig/pair-sq.tex
\begin{minipage}[t]{0.44\columnwidth}
\begin{center}
\begin{tikzpicture}[rounded corners=1pt,scale=0.7,baseline=(in-cons-1)]

\node[trace] at (1,0)
(app-1)          {\app{f}};
\node[valueIn,left of=app-1]
(in-cons-1)      {\ttt{:}};
\node[valueIn,below left of=in-cons-1]
(in-el-1)        {\ttt{1}};
\node[value,right of=app-1]
(out-cons-1)     {\ttt{:}};
\node[value,below left of=out-cons-1]
(out-el-1)       {\ttt{1}};

\drawArg{app-1.west}{in-cons-1.east}
\draw[subvalue] (in-cons-1.south west) -- (in-el-1.north east);
\draw[val] (app-1.east) -- (out-cons-1.west);
\draw[subvalue] (out-cons-1.south west) -- (out-el-1.north east);

\node[trace,below of=out-cons-1,xshift=-3mm,yshift=-4mm]
(app-2)          {\app{f}};
\node[valueIn,left of=app-2]
(in-cons-2)      {\ttt{:}};
\node[valueIn,below left of=in-cons-2]
(in-el-2)        {\ttt{3}};
\node[value,right of=app-2]
(out-cons-2)     {\ttt{:}};
\node[value,below left of=out-cons-2]
(out-el-2)       {\ttt{3}};

\draw[subvalue] (in-cons-1.south east) -- (in-cons-2.north west);
\drawArg{app-2.west}{in-cons-2.east}
\draw[subvalue] (in-cons-2.south west) -- (in-el-2.north east);
\draw[subvalue] (out-cons-1.south) -- (app-2.north);
\draw[val] (app-2.east) -- (out-cons-2.west);
\draw[subvalue] (out-cons-2.south west) -- (out-el-2.north east);

\node[trace,below of=out-cons-2,xshift=-3mm,yshift=-4mm]
(app-3)          {\app{f}};
\node[valueIn,left of=app-3]
(in-nil)         {\ttt{[]}};
\node[value,right of=app-3]
(out-nil)        {\ttt{[]}};

\draw[subvalue] (in-cons-2.south east) -- (in-nil.north west);
\drawArg{app-3.west}{in-nil.east}
\draw[subvalue] (out-cons-2.south) -- (app-3.north);
\draw[val] (app-3.east) -- (out-nil.west);

\end{tikzpicture}
\end{center}
\end{minipage}%
\begin{minipage}[t]{0.49\columnwidth}
\begin{center}
\begin{tikzpicture}[rounded corners=1pt,scale=0.7,baseline=(in-cons-1)]

\node[valueIn] at (0,0)
(in-cons-1)      {\ttt{:}};
\node[trace,right of=in-cons-1]
(app-1)          {\app{f}};
\node[valueIn,below left of=in-cons-1]
(in-el-1)        {\ttt{1}};
\node[value,right of=app-1]
(out-cons-1)     {\ttt{:}};
\node[valueNew,below left of=out-cons-1,xshift=-2mm]
(out-el-1)       {\ttt{(,)}};
\node[value,below left of=out-el-1,yshift=-2mm]
(out-pairl-1)    {\ttt{1}};
\node[traceNew,right of=out-pairl-1]
(app-sq-1)       {\app{sq}};
\node[valueNew,right of=app-sq-1]
(out-sq-1)       {\ttt{1}};

\drawArg{app-1.west}{in-cons-1.east}
\draw[subvalue] (in-cons-1.south west) -- (in-el-1.north east);
\draw[val] (app-1.east) -- (out-cons-1.west);
\draw[subvalueChanged] (out-cons-1.south west) -- (out-el-1.north east);
\drawArgNew{app-sq-1.west}{out-pairl-1.east}
\draw[valNew] (app-sq-1.east) -- (out-sq-1.west);
\draw[subvalueNew] (out-el-1.south) -- (out-pairl-1.north east);
\draw[subvalueNew] (out-el-1.south) -- (app-sq-1.north west);

\node[valueIn,below right of=in-cons-1,yshift=-16mm]
(in-cons-2)      {\ttt{:}};
\node[trace,right of=in-cons-2]
(app-2)          {\app{f}};
\node[valueIn,below left of=in-cons-2]
(in-el-2)        {\ttt{3}};
\node[value,right of=app-2]
(out-cons-2)     {\ttt{:}};
\node[valueNew,below left of=out-cons-2,xshift=-2mm]
(out-el-2)       {\ttt{(,)}};
\node[value,below left of=out-el-2,yshift=-2mm]
(out-pairl-2)    {\ttt{3}};
\node[traceNew,right of=out-pairl-2]
(app-sq-2)       {\app{sq}};
\node[valueNew,right of=app-sq-2]
(out-sq-2)       {\ttt{9}};

\draw[subvalue] (in-cons-1.south east) -- (in-cons-2.north west);
\drawArg{app-2.west}{in-cons-2.east}
\draw[subvalue] (in-cons-2.south west) -- (in-el-2.north east);
\draw[subvalue] (out-cons-1.south) -- (app-2.north);
\draw[val] (app-2.east) -- (out-cons-2.west);
\draw[subvalueChanged] (out-cons-2.south west) -- (out-el-2.north east);
\drawArgNew{app-sq-2.west}{out-pairl-2.east}
\draw[valNew] (app-sq-2.east) -- (out-sq-2.west);
\draw[subvalueNew] (out-el-2.south) -- (out-pairl-2.north east);
\draw[subvalueNew] (out-el-2.south) -- (app-sq-2.north west);

\node[valueIn,below right of=in-cons-2,yshift=-16mm]
(in-nil)         {\ttt{[]}};
\node[trace,right of=in-nil]
(app-3)          {\app{f}};
\node[value,right of=app-3]
(out-nil)        {\ttt{[]}};

\draw[subvalue] (in-cons-2.south east) -- (in-nil.north west);
\drawArg{app-3.west}{in-nil.east}
\draw[subvalue] (out-cons-2.south) -- (app-3.north);
\draw[val] (app-3.east) -- (out-nil.west);

\end{tikzpicture}
\end{center}
\end{minipage}

%% file: page-2.tex
\subsection*{Executing changes via changes in executions}

Here is a rough sketch of the idea more formally. Suppose we already have a
baseline big-step evaluation relation $\eval_0$ for our language. We start by
defining a relation $e \eval v$ for programs represented as term graphs,
consistent with the baseline semantics in that $e \eval v$ implies $|e| \eval_0
|v|$ where $|e|$ is the ``tree-unravelling'' \cite{ehrig99} of $e$. Then the
idea would be to derive an \emph{incremental} evaluation relation $\deltaeval$
satisfying: \begin{equation} e \eval v \textcolor{Gray}{\text{ and }} e, \delta
e \deltaeval \delta v \textcolor{Gray}{\text{ implies }} e \oplus \delta e \eval
v \oplus \delta v
\end{equation}

\noindent where $\oplus$ applies a delta to a term graph. Note that this is
quite different from the static differentiation problem \cite{cai14,elliott17},
since $\delta e$ describes an arbitrary change to $e$.

Now, in our setting, it is always possible to compute the difference $v \ominus
\oneprime{v}$ between two outputs $v$ and $\oneprime{v}$, or more generally the
difference $D \ominus \oneprime{D}$ between two executions $D: e \eval v$ and
$\oneprime{D}: \oneprime{e} \eval \oneprime{v}$, simply by comparing the
term-graph representations \cite{perera13}. So in fact we can always trivially
exhibit a relation satisfying property (1) by defining $e, \delta e
\eval_{\delta} \oneprime{v} \ominus v$ where $e \eval v$ and $e \oplus \delta e
\eval v'$, supposing that $\ominus$ satisfies $v \oplus (v' \ominus v) = v'$
\cite{cai14}.

We are left with at least two significant challenges, however. First, we must
take care to balance precision with the integrity constraints of the term graph.
We must define $e \eval v$ so that distinct nodes are always used to represent
distinct terms (so $v$ remains a well-formed term graph), but also so that nodes
are reused systematically in order to obtain precise deltas.

As an example of how nodes can be safely reused, suppose we extracted the
expression \kw{(x, sq x)} in the definition of \kw{f} into a new function
\kw{g}, replacing free occurrences of \kw{x} by uses of the argument \kw{y} of
the new function \cite{fowler99}:

\begin{graylstlisting}
f [] = []
f (x : xs) = $\New{g}$ x : f xs

$\New{g y =}$ ($\New{y}$, sq $\New{y}$)
\end{graylstlisting} %

\noindent Such refactorings are common programming ``micro-steps''
\cite{perera04}. They clearly have no effect on the output $v$ considered as its
tree-unravelling $|v|$. But we might also observe that although the expression
responsible for constructing the pairs is now inside \kw{g}, the values used to
close that expression are the same as before. Moreover \kw{g} simply delegates
to that expression. So it should be safe to use the same nodes for the pairs in
the new execution as we did in the previous one, and view the change simply as a
``rewiring'' of the computation:

\input{fig/extract-g}

\vspace{3pt}
On the other hand, given that \kw{f} has now simplified to \kw{map g},
suppose we adjusted our test case as follows:

\begin{graylstlisting}
   $\New{map g}$ [1, 3] => $\New{[}$(1, 1)$\New{,}$ (3, 9)$\New{]}$
\end{graylstlisting}

\noindent This is also a refactoring, and so again the output unravels to the
same tree as before. But now a different expression (one found in the body of
\kw{map}) is responsible for constructing the list nodes in the output. It would
therefore be unsafe to use the same nodes as we did previously to construct the
list, at least not without additional global knowledge (for example that \kw{f}
is not called anywhere). This explains the new lists nodes that were highlighted
in the output delta above:

\input{fig/map-g}

\vspace{3pt}
A second challenge is to efficiently implement $\eval_{\delta}$ so that
\emph{memoised} \cite{michie68} parts of a previous evaluation $D: e \eval v$
can be reused to produce $D': e \oplus \delta e \eval v'$. This leads naturally
to the question of whether existing approaches to incremental computation, in
particular \emph{self-adjusting computation} and related work
\cite{acar05,hammer14,hammer15}, can be be brought to bear on this problem, by
using a general-purpose incremental language as a metalanguage to derive an
interpreter able to respond to program changes. While \citet{hammer15} show that
such an approach is possible, this technique does not in itself give rise to a
direct program-change semantics for the object language, but rather one which is
implicit in the (complex) incremental behaviour of the metalanguage.

\vspace{3pt}
\subsection*{From programs to programming}

Programming is the activity of modifying a program in order to bring about
specific changes in its behaviour. Yet programming language theory almost
exclusively focuses on the meaning of \emph{programs} \cite{petricek19}. We
outlined a ``change-oriented'' viewpoint from which the meaning of a program
change is a change to the program's meaning. In so doing we skipped over many
difficult questions.

We raise one question now with a view to future discussion. Can we extend the
change-oriented perspective to the \emph{static} meaning of a program, such as
its type? For example, \citet{omar17} define a type system which incrementally
maintains a proof that a program being edited is well-typed. Viewing such a
system as generating a higher-order witness (explaining how specific edits
induce specific changes to the type of the program) suggests a static version of
our principle: the \emph{type of a program change} is a \emph{change in the
program's type}. Equipping \emph{changes} with types may improve our ability to
produce useful error messages, especially if we can extend type deltas to
structural reorganisations like the ones considered here.

%% file: fig/extract-g.tex
\begin{center}
\begin{tikzpicture}[rounded corners=1pt,scale=0.7,baseline=(in-cons-1)]

\node[valueIn] at (0,0)
(in-cons-1)      {\ttt{:}};
\node[trace,right of=in-cons-1]
(app-1)          {\app{f}};
\node[valueIn,below left of=in-cons-1]
(in-el-1)        {\ttt{1}};
\node[value,right of=app-1]
(out-cons-1)     {\ttt{:}};
\node[traceNew,below left of=out-cons-1]
(out-app-1)       {\app{g}};
\node[value,right of=out-app-1]
(out-el-1)       {\ttt{(,)}};
\node[value,below left of=out-el-1,yshift=-2mm]
(out-pairl-1)    {\ttt{1}};
\node[trace,right of=out-pairl-1]
(app-sq-1)       {\app{sq}};
\node[value,right of=app-sq-1]
(out-sq-1)       {\ttt{1}};

\drawArg{app-1.west}{in-cons-1.east}
\draw[subvalue] (in-cons-1.south west) -- (in-el-1.north east);
\draw[val] (app-1.east) -- (out-cons-1.west);
\draw[subvalueChanged] (out-cons-1.south west) -- (out-app-1.north east);
\drawArgNew{out-app-1.west}{in-el-1.east}
\draw[valNew] (out-app-1.east) -- (out-el-1.west);
\drawArg{app-sq-1.west}{out-pairl-1.east}
\draw[val] (app-sq-1.east) -- (out-sq-1.west);
\draw[subvalue] (out-el-1.south) -- (out-pairl-1.north east);
\draw[subvalue] (out-el-1.south) -- (app-sq-1.north west);

\node[valueIn,below right of=in-cons-1,yshift=-16mm]
(in-cons-2)      {\ttt{:}};
\node[trace,right of=in-cons-2]
(app-2)          {\app{f}};
\node[valueIn,below left of=in-cons-2]
(in-el-2)        {\ttt{3}};
\node[value,right of=app-2]
(out-cons-2)     {\ttt{:}};
\node[traceNew,below left of=out-cons-2]
(out-app-2)       {\app{g}};
\node[value,right of=out-app-2]
(out-el-2)       {\ttt{(,)}};
\node[value,below left of=out-el-2,yshift=-2mm]
(out-pairl-2)    {\ttt{3}};
\node[trace,right of=out-pairl-2]
(app-sq-2)       {\app{sq}};
\node[value,right of=app-sq-2]
(out-sq-2)       {\ttt{9}};

\draw[subvalue] (in-cons-1.south east) -- (in-cons-2.north west);
\drawArg{app-2.west}{in-cons-2.east}
\draw[subvalue] (in-cons-2.south west) -- (in-el-2.north east);
\draw[subvalue] (out-cons-1.south) -- (app-2.north);
\draw[val] (app-2.east) -- (out-cons-2.west);
\draw[subvalueChanged] (out-cons-2.south west) -- (out-app-2.north east);
\drawArgNew{out-app-2.west}{in-el-2.east}
\draw[valNew] (out-app-2.east) -- (out-el-2.west);
\drawArg{app-sq-2.west}{out-pairl-2.east}
\draw[val] (app-sq-2.east) -- (out-sq-2.west);
\draw[subvalue] (out-el-2.south) -- (out-pairl-2.north east);
\draw[subvalue] (out-el-2.south) -- (app-sq-2.north west);

\node[valueIn,below right of=in-cons-2,yshift=-16mm]
(in-nil)         {\ttt{[]}};
\node[trace,right of=in-nil]
(app-3)          {\app{f}};
\node[value,right of=app-3]
(out-nil)        {\ttt{[]}};

\draw[subvalue] (in-cons-2.south east) -- (in-nil.north west);
\drawArg{app-3.west}{in-nil.east}
\draw[subvalue] (out-cons-2.south) -- (app-3.north);
\draw[val] (app-3.east) -- (out-nil.west);

\end{tikzpicture}
\end{center}

%% file: fig/map-g.tex
\begin{center}
\begin{tikzpicture}[rounded corners=1pt,scale=0.7,baseline=(in-cons-1)]

\node[valueIn] at (0,0)
(in-cons-1)      {\ttt{:}};
\node[traceNew,right of=in-cons-1]
(app-1)          {\app{map g}};
\node[valueIn,below left of=in-cons-1]
(in-el-1)        {\ttt{1}};
\node[valueNew,right of=app-1]
(out-cons-1)     {\ttt{:}};
\node[trace,below left of=out-cons-1]
(out-app-1)       {\app{g}};
\node[value,right of=out-app-1]
(out-el-1)       {\ttt{(,)}};
\node[value,below left of=out-el-1,yshift=-2mm]
(out-pairl-1)    {\ttt{1}};
\node[trace,right of=out-pairl-1]
(app-sq-1)       {\app{sq}};
\node[value,right of=app-sq-1]
(out-sq-1)       {\ttt{1}};

\drawArgNew{app-1.west}{in-cons-1.east}
\draw[subvalue] (in-cons-1.south west) -- (in-el-1.north east);
\draw[valNew] (app-1.east) -- (out-cons-1.west);
\draw[subvalueNew] (out-cons-1.south west) -- (out-app-1.north east);
\drawArg{out-app-1.west}{in-el-1.east}
\draw[val] (out-app-1.east) -- (out-el-1.west);
\drawArg{app-sq-1.west}{out-pairl-1.east}
\draw[val] (app-sq-1.east) -- (out-sq-1.west);
\draw[subvalue] (out-el-1.south) -- (out-pairl-1.north east);
\draw[subvalue] (out-el-1.south) -- (app-sq-1.north west);

\node[valueIn,below right of=in-cons-1,yshift=-16mm]
(in-cons-2)      {\ttt{:}};
\node[traceNew,right of=in-cons-2]
(app-2)          {\app{map g}};
\node[valueIn,below left of=in-cons-2]
(in-el-2)        {\ttt{3}};
\node[valueNew,right of=app-2]
(out-cons-2)     {\ttt{:}};
\node[trace,below left of=out-cons-2]
(out-app-2)       {\app{g}};
\node[value,right of=out-app-2]
(out-el-2)       {\ttt{(,)}};
\node[value,below left of=out-el-2,yshift=-2mm]
(out-pairl-2)    {\ttt{3}};
\node[trace,right of=out-pairl-2]
(app-sq-2)       {\app{sq}};
\node[value,right of=app-sq-2]
(out-sq-2)       {\ttt{9}};

\draw[subvalue] (in-cons-1.south east) -- (in-cons-2.north west);
\drawArgNew{app-2.west}{in-cons-2.east}
\draw[subvalue] (in-cons-2.south west) -- (in-el-2.north east);
\draw[subvalueNew] (out-cons-1.south) -- (app-2.north);
\draw[valNew] (app-2.east) -- (out-cons-2.west);
\draw[subvalueNew] (out-cons-2.south west) -- (out-app-2.north east);
\drawArg{out-app-2.west}{in-el-2.east}
\draw[val] (out-app-2.east) -- (out-el-2.west);
\drawArg{app-sq-2.west}{out-pairl-2.east}
\draw[val] (app-sq-2.east) -- (out-sq-2.west);
\draw[subvalue] (out-el-2.south) -- (out-pairl-2.north east);
\draw[subvalue] (out-el-2.south) -- (app-sq-2.north west);

\node[valueIn,below right of=in-cons-2,yshift=-16mm]
(in-nil)         {\ttt{[]}};
\node[traceNew,right of=in-nil]
(app-3)          {\app{map g}};
\node[valueNew,right of=app-3]
(out-nil)        {\ttt{[]}};

\draw[subvalue] (in-cons-2.south east) -- (in-nil.north west);
\drawArgNew{app-3.west}{in-nil.east}
\draw[subvalueNew] (out-cons-2.south) -- (app-3.north);
\draw[valNew] (app-3.east) -- (out-nil.west);

\end{tikzpicture}
\end{center}